\documentclass[acmsmall,screen,authorversion]{acmart}

\usepackage{amsmath}
\usepackage{subcaption}
\usepackage{supertabular}
\usepackage{stmaryrd}
\usepackage{xcolor}
\usepackage{multirow}
\usepackage{calc}
\usepackage[T1]{fontenc}
\usepackage{graphicx}
\usepackage{xspace}
\usepackage{dashrule}  
\usepackage{stackrel}
\usepackage{enumerate}
\usepackage{framed}
\usepackage{bussproofs}
\usepackage{mdframed}  
\usepackage{comment}
\usepackage{ifthen}
\usepackage{pdftexcmds}
\usepackage{mathpartir}
\usepackage{fancyvrb}
\usepackage{booktabs}
\usepackage{tikz}
\usepackage{listings}
\usepackage{url}

\usepackage{lstparams}
\usepackage{lstcoq}

\newcommand{\codeplus}[3]{%
  \lstinputlisting[%
        #3,
        linerange={#2},
        rangebeginprefix=(*\ begin\ ,%
        rangebeginsuffix=\ *),%
        rangeendprefix=(*\ end\ ,%
        rangeendsuffix=\ *),
        includerangemarker=false]{#1}}

\usepackage[capitalize, nameinlink]{cleveref}

\usepackage{siunitx}
\newcommand\cd[1]{\lstinline[breakatwhitespace]{#1}}
\newcommand\cdh[1]{\lstinline[language=Haskell,breakatwhitespace]{#1}}

\newcommand\cdc[1]{\lstinline[language=Coq,breakatwhitespace]{#1}}

\newcommand\hstocoq{\texttt{hs-to-coq}\xspace}

\newcommand\ie{\textit{i.e.,\ }}
\newcommand\eg{\textit{e.g.,\ }}
\newcommand{\LiquidHaskell}{\textsc{Liquid Haskell}\xspace}
\newcommand{\optim}{optimistic}
\newcommand{\pessim}{pessimistic}
\newcommand{\Optim}{Optimistic}
\newcommand{\Pessim}{Pessimistic}

\newcommand\unitty{\textsc{unit}}
\newcommand\listty{\textsc{list}}
\newcommand\letsym{\textsc{let}}
\newcommand\lettm[3]{\letsym\;#1\;=\;#2\;\textsc{in}\;#3}
\newcommand\niltm{\textsc{Nil}}
\newcommand\constm{\textsc{Cons}}
\newcommand\foldsym{\textsc{foldr}}
\newcommand\foldtm[3]{\foldsym\;#1\;#2\;#3}

\newcommand*\ccbv[5]{#1,#2 \Downarrow^{\mathrm{CV}}_{#3} #4,#5}

\setcopyright{rightsretained}
\acmPrice{}
\acmDOI{10.1145/3473585}
\acmYear{2021}
\copyrightyear{2021}
\acmJournal{PACMPL}
\acmVolume{5}
\acmNumber{ICFP}
\acmArticle{80}
\acmMonth{8}

\begin{document}
\title{Reasoning about the Garden of Forking Paths}

\author{Yao Li}
\email{liyao@cis.upenn.edu}
\orcid{0000-0001-8720-883X}
\affiliation{%
  \institution{University of Pennsylvania}
  \streetaddress{3330 Walnut St}
  \city{Philadelphia}
  \state{PA}
  \postcode{19104}
  \country{USA}
}

\author{Li-yao Xia}
\email{xialiyao@cis.upenn.edu}
\orcid{0000-0003-2673-4400}
\affiliation{%
  \institution{University of Pennsylvania}
  \streetaddress{3330 Walnut St}
  \city{Philadelphia}
  \state{PA}
  \postcode{19104}
  \country{USA}
}

\author{Stephanie Weirich}
\email{sweirich@cis.upenn.edu}
\orcid{0000-0002-6756-9168}
\affiliation{%
  \institution{University of Pennsylvania}
  \streetaddress{3330 Walnut St}
  \city{Philadelphia}
  \state{PA}
  \postcode{19104}
  \country{USA}
}

\begin{abstract}
  Lazy evaluation is a powerful tool for functional programmers. It enables the
  concise expression of on-demand computation and a form of compositionality not
  available under other evaluation strategies. However, the stateful nature of
  lazy evaluation makes it hard to analyze a program's computational cost,
  either informally or formally. In this work, we present a novel and simple
  framework for formally reasoning about lazy computation costs based on a
  recent model of lazy evaluation: clairvoyant call-by-value. The key feature of
  our framework is its simplicity, as expressed by our definition of
  the \emph{clairvoyance monad}. This monad is both simple to define~(around 20
  lines of Coq) and simple to reason about. We show that this monad can be
  effectively used to mechanically reason about the computational cost of lazy
  functional programs written in Coq.
\end{abstract}



\maketitle

\bibliographystyle{ACM-Reference-Format}
\citestyle{acmauthoryear}   

\section{Introduction}\label{sec:intro}

Lazy evaluation~\citep{lazy}, or the \emph{call-by-need} calling convention, is
a distinguishing feature in some functional programming languages---Haskell
being the most notable example.
Rather than evaluating eagerly, a lazy evaluator stores computations in a
thunk and only evaluates the thunk when the data is needed. This feature
avoids unneeded computation and enables better
modularity in functional programming~\citep{fp-matters}.
However, with convenience in expressiveness comes challenge in
reasoning---especially so for cost analysis---because it's far less
obvious if, when, and how much a computation is evaluated.
We believe proof assistants can help to reason formally about semantics so
intricate and subtle.

However, modeling laziness formally is difficult.
The semantics for call-by-need evaluation is more complex than that of
call-by-name or call-by-value, which can be described merely through
substitution. Traditional presentations~\cite{launchbury1993,josephs1989} of
call-by-need semantics are fundamentally stateful, based on heaps that
contain thunks which must be updated during evaluation.

Rather than directly dealing with such complexity, we take advantage of a new
way of modeling call-by-need: \emph{clairvoyant
call-by-value}~\citep{clairvoyant}. The key observation of this new model is
that although \emph{whether} a term gets evaluated matters, it doesn't
matter \emph{when} in run-time cost analysis. Therefore, instead of storing the
computations in a thunk, the clairvoyant call-by-value model makes use of
nondeterminism to evaluate the data in one branch and skip evaluation in
another. Eventually, one successful branch of evaluation will faithfully model
the result and cost of the call-by-need evaluation.

Based on clairvoyant evaluation, we propose a novel framework for reasoning
about laziness using the Coq proof assistant, a dependently-typed interactive
theorem prover~\citep{coq}. Our framework is based on an annotated model similar
to that of \citet{danielsson-08} and of \citet{liquidate}, but our work does not
require an explicit notion of laziness to reason about computation costs under
lazy evaluation.

We make the following contributions:
\begin{itemize}
  \item We present the clairvoyance monad, a model of laziness that can be
    shallowly embedded in Coq by distilling two core features of clairvoyant
    evaluation: nondeterminism and cost accumulation. One key feature of the
    clairvoyance monad is its simplicity: it consists of only a handful of basic
    combinators and can be defined with around 20 lines of code in
    Coq~(\Cref{sec:clairvoyance-monad}).

  \item We develop a translation from a lazy calculus to programs in this
    monad that captures the semantics of clairvoyant call-by-value evaluation.
    Compared with the denotational semantics of \citet{clairvoyant}, our
    translation deals with typed programs, does not rely on domain
    theory, and accounts for the cost of every nondeterministic
    execution~(\Cref{sec:translation}).

  \item One challenge arising from clairvoyant evaluation is to reason about
    nondeterminism. We develop dual logics of \emph{over-}
    and \emph{under-approximations} similar to those
    of \citet{hoare-logic,rev-hoare,incorrectness} to reason about lazy
    programs. We show that formal reasoning of computation cost based on these
    logics can be done in a \emph{local} and \emph{modular}
    way~(\Cref{sec:reasoning}).

  \item We demonstrate the usefulness of our technique via case studies of tail
    recursion that capture the unique characteristics of lazy
    evaluation~(\Cref{sec:case-study}).
\end{itemize}

Our paper ends with a comparison of related research in \cref{sec:related-work} and a
discussion of future work in \cref{sec:conclusion}.
In the next section, below, we introduce our approach at a high level and present our running example.

\section{Motivating example}\label{sec:example}

We start by providing an informal overview of our approach on a small example
that exhibits laziness. Consider the following program, written in Gallina, the
functional programming language of the Coq proof assistant~\cite{coq}:
\codeplus{example.v}{pcomp}{language=Coq}
Gallina can be compiled to use lazy evaluation via extraction to Haskell. The
functions \cdc{append} and \cdc{take} in this example are equivalent to their
Haskell counterparts, and their definitions are shown
in \cref{fig:source-example}.~\footnote{For reasons that we will explain
in \cref{sec:translation}, these functions are written in A-Normal
Form~\cite{anf}, but that doesn't matter so much here.}  These examples use
Gallina's inductively defined lists, which are a subset of Haskell's list type.
Although working with infinite data types is another useful application of lazy
evaluation, many algorithms manipulate only finite data
structures~\cite{purely-functional}. Hence, we believe inductive lists are
representative of how lists are used in practice even in Haskell.

\begin{figure}[t]
\codeplus{example.v}{fig1}{language=Coq}
\caption{The pure functional definitions of \cdc{append} and \cdc{take}.}\label{fig:source-example}
\end{figure}

To estimate the time it takes to evaluate a program, its cost, we can start by
counting the number of steps in some operational semantics, or some
proportional quantity. Let us count function calls informally.

Lazy evaluation leads us immediately to an impasse, because it is not even clear
what it means to ``run'' a lazy program. Lazy programs are demand-driven, so we
have to specify some model of ``demand''. A common working model is that lazy
programs will be forced during the evaluation of a whole program, but it is
not so practical to reason about the behaviors of arbitrary programs.  A more useful
approach is to start from a more familiar place: call-by-value. Indeed, programs
under the call-by-value evaluation strategy have a relatively straightforward
cost model. Laziness adds a twist to it: we might not need all of the result,
in which case we allow ourselves to skip some computations.

With that new ability, we face the problem of
deciding \emph{which} computations to skip. This decision inherently
depends on how much of the overall result will be needed.
For concreteness, let us require all of our example list
\cdc{take n (append xs ys)} to be evaluated.
We start by evaluating \cdc{append xs ys}, unfolding the
program in call-by-value. There are two cases to consider: the length of
\cdc{xs} may be less than \cdc{n}, in which case we will fully evaluate
\cdc{append xs ys} in \cdc{length xs + 1} calls---where the final \cdc{nil}
takes one call. Or \cdc{length xs} may be greater than or equal to \cdc{n},
then we can stop after \cdc{n} calls, leaving the result of the next call
``undefined''. Either way, we will produce some partially defined list \cdc{zs}
after at most \cdc{n} calls. We then let \cdc{take n zs} run to the end in at
most \cdc{n + 1} calls, thus producing all elements of \cdc{take n (append xs
ys)}, as we demanded initially.  In total, that took at most \cdc{2 * n + 1}
calls. In particular, that cost is independent of the length of \cdc{xs} or
\cdc{ys}.
That exemplifies one of the core motivations of laziness: you only pay for what
you need.

That idea of ``call-by-value with a twist'' is made formal by the concept of
\emph{clairvoyant evaluation}~\cite{clairvoyant}.

\subsection{Clairvoyant Evaluation}

The key to formalize the reasoning above is to view lazy programs
as \emph{nondeterministic} programs. Clairvoyant evaluation works in a way
similar to nondeterministic automata, which choose one of multiple successor
states by \emph{``guessing''} the path to success. In our earlier reasoning, we
evaluated \cdc{append xs ys} in call-by-value until we decided to stop at a
point. \emph{When} to stop was not decided by the state of the program, but by a
``guess'' based on the clairvoyant knowledge that we would only need \cdc{n}
elements in the end. This intuition allows us to define a general semantics that
the meaning of a lazy program comprises all of its nondeterministic evaluations
and the meaning can be refined later in light of new external information.

The equivalence of clairvoyant evaluation to the natural heap-based
definition of laziness~\cite{launchbury1993} was proved by~\citet{clairvoyant}:
the cost of any execution in clairvoyant call-by-value is an upper bound of the
cost in call-by-need, and there is some clairvoyant execution whose
cost is actually the same as in call-by-need.
In this paper, we carry on with the clairvoyant interpretation.

Taking the \cdc{append} function as an example~(recall its definition
in \cref{fig:source-example}), when it makes a recursive call, we fork
the evaluation into two branches: in branch~(1), we perform the recursive
call; and in branch~(2), we skip that call. A skipped call yields a placeholder
value as a result, which we call $\bot$ or \cdc{Undefined}.

\begin{figure}[t]
\includegraphics[page=1,width=0.8\textwidth]{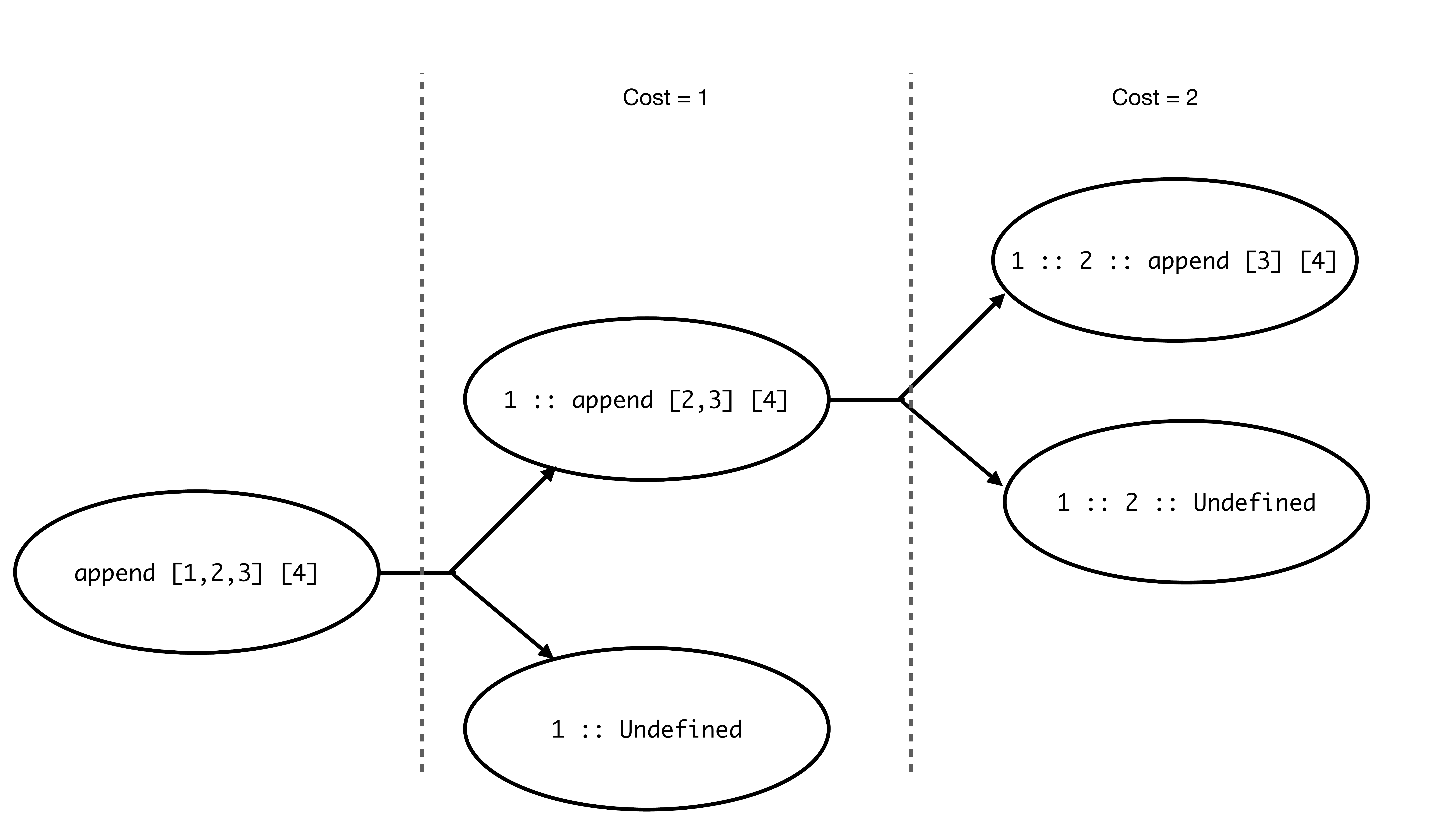}
\caption{Parts of the nondeterministic clairvoyant call-by-value evaluation of
   the \cdc{append} function applied to two lists \cdc{[1, 2, 3]}
   and \cdc{[4]}.}\label{fig:clairvoyant-branches}
\end{figure}

Suppose that all future demands only require the first element of the result
list, then branch~(2) would suffice for offering that result. However, if a
future demand requests more than that, branch~(2) would fail to proceed because
the requested data is \cdc{Undefined}. Therefore, branch~(2) would get stuck and
not yield any result at all. Fortunately, there would still be branch~(1) to
return the result. Furthermore, in branch~(1), the \cdc{append} function may
make another recursive call to itself, as long as the first argument list is not
\cdc{nil}. In that case, this branch would be once again forked into two
branches. This is illustrated by \cref{fig:clairvoyant-branches}.

Although lazy programs are now interpreted nondeterministically, nondeterminism
is used in a very controlled manner.  The only choices we make are whether to
perform a computation or to skip it.  This means that the value of a program, if
it exists, is still unique in some sense: the only possible changes are that
parts of the value are replaced with \cdc{Undefined}.

If we know which branch leads to a successful evaluation for the \cdc{append}
function, we can just look at that branch and add its cost to obtain the total
cost of the program, which gives us a local reasoning methodology. Of course we
cannot know this in advance, but there are some reasoning principles that can
help.

\subsection{A Dual Reasoning Principle}\label{subsec:example-spec}

We would like to have a reasoning methodology that is both \emph{local}
and \emph{modular}, just like what we would expect from functional
programming. This means that we should be able to use some relations to specify
the behaviors of each individual function. And when we want to reason about a
program, we can just do that by composing the relations of its functions.

We use a dual reasoning principle to achieve the locality and modularity for
clairvoyant call-by-value evaluation. First, we have a \emph{\pessim{}}
specification that describes the behaviors of \emph{all} of the function's
nondeterministic branches. The \pessim{} specification can offer us an accurate
description of functional correctness under call-by-need evaluation. However,
the specification is \pessim{} because it does not rule out the branches that
contain redundant steps and would not appear in an actual call-by-need
evaluation.

To be more selective in those branches, we use an \optim{} specification
that describes the behaviors on a specific branch. The specification is \optim{}
because it can be used to specify a more accurate bound for the cost under
call-by-need evaluation.

\begin{figure}[t]
\includegraphics[page=2,width=0.8\textwidth]{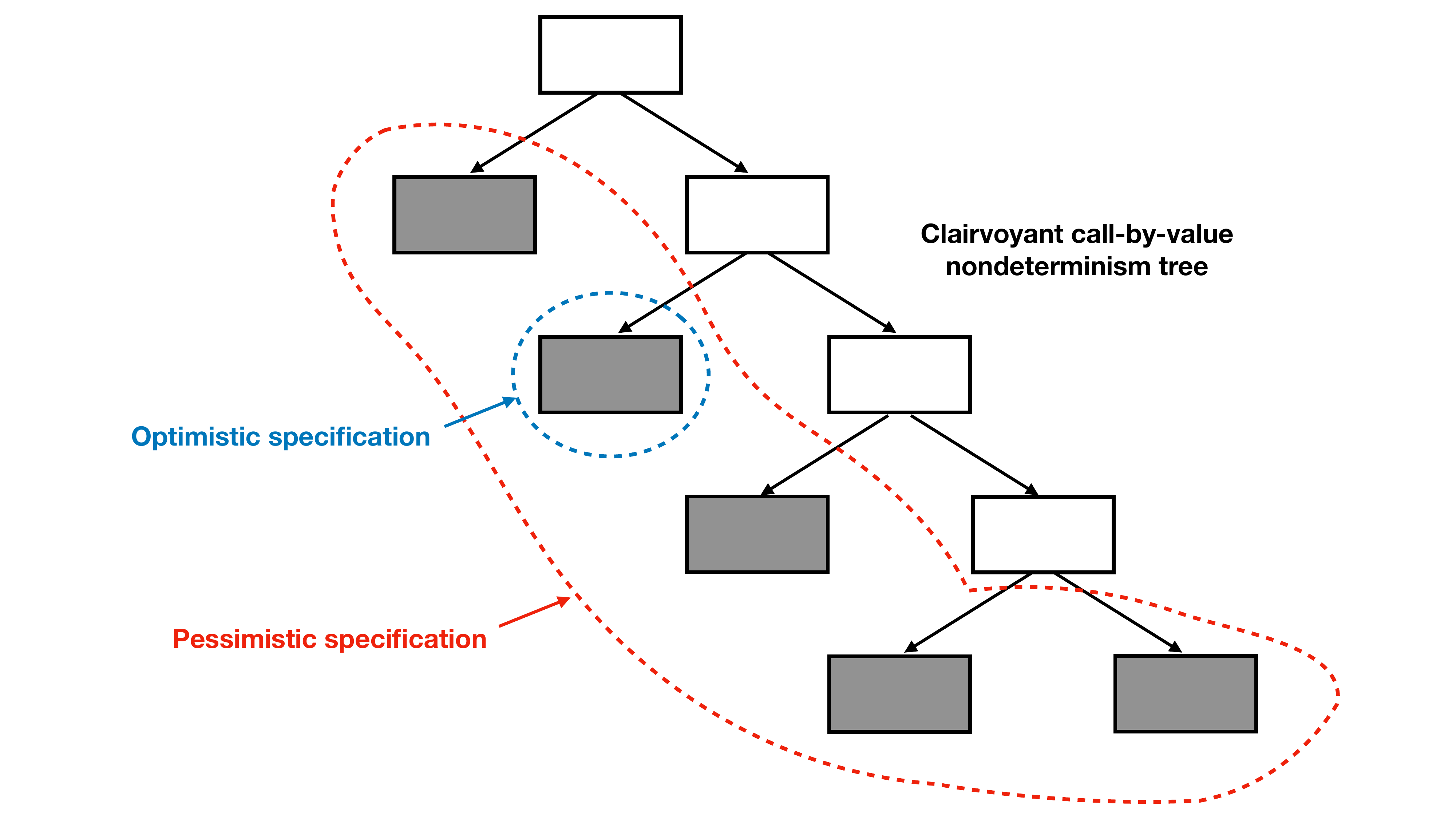}
\caption{The relations among a clairvoyant evaluation, a \pessim{}
   specification, and an \optim{} specification. The tree in the middle of the
   figure represents the nondeterminism tree of clairvoyant evaluation. The gray
   nodes represent the end results of their branches. A \pessim{} specification
   specifies the nodes in the red circle. And an \optim{} specification
   specifies the node in the blue circle.}\label{fig:optim-and-pessim}
\end{figure}

\Cref{fig:optim-and-pessim} shows the relations among a clairvoyant evaluation,
a \pessim{} specification, and an \optim{} specification.

Getting back to \cdc{append}, its \pessim{} specification states:
\begin{quote}
For \emph{all} the nondeterministic branches of the \cdc{append} function, if
the branch evaluates successfully, it will return a cost $c \in [1, \text{length
} xs + 1]$.
\end{quote}
For simplicity, we omit the functional correctness part of the specification
here. The \pessim{} specification only specifies a coarse range
for \cdc{append}'s costs. If we want to reason about our earlier
example \cdc{p}, we could not deduce that its cost would never go over \cdc{n}
by reasoning about \cdc{append} and \cdc{take} abstractly using their
specifications. Instead, we can only deduce that the upper bound of the cost is
the length of \cdc{xs}, which may be much larger than \cdc{n}.

For this sort of analysis, we need the \optim{} specification of \cdc{append}:
\begin{quote}
For any number $n \in [1, \text{length }xs]$ (or $n \in [1, \text{length }xs +
1]$ if $xs$ does not contain any undefined part),~\footnote{When $xs$ does not
contain any undefined part, \cdc{append} might go over the entire list and take
one extra cost pattern matching on \cdc{nil}. This is the only case the cost
of \cdc{append} will be bigger than the length of $xs$.} there \emph{exists} a nondeterministic branch
of the \cdc{append} function that evaluates successfully and returns a cost $c =
n$.
\end{quote}
A major difference between the \pessim{} and the \optim{} specifications is that
the latter does not only show a range of costs; it shows what exactly are the
possible costs within this range. With the help of this specification, we can
``pick'' one branch where the possible cost, which might be much smaller than
the length of $xs$, barely suffices for producing a list that \cdc{take} needs.

Both the \pessim{} and \optim{} specifications can be proved on the \cdc{append}
function. And when reasoning about a larger program like \cdc{p}, all we need is
to compose the specifications of \cdc{append} with the specifications of other
functions like \cdc{take}.

\subsection{The Missing Pieces}

So far, we have informally discussed our methodology. The main missing piece to
develop in the rest of the paper is to implement this methodology in the formal
environment of the Coq proof assistant.

The first step is to associate the pure functional programs with versions that
track execution costs and allow nondeterminism. It turns out that we can do that
with monads~\citep{moggi-monad, wadler-monad}. In the next section, we define the
\emph{clairvoyance monad}, which distills the main features of clairvoyant
evaluation. One attraction of the clairvoyance monad is its simplicity: its
core definitions consist of merely 21 nonblank, noncomment lines of code in
Coq. Then, we translate pure functions into the clairvoyance monad in~\cref{sec:translation}.
The final step, in \cref{sec:reasoning}, is to build a program logic for the
clairvoyance monad that enables local and modular formal cost analysis in the
style of \optim{} and \pessim{} specifications.

\section{The clairvoyance monad}\label{sec:clairvoyance-monad}

The clairvoyance monad (\cref{fig:clairvoyance-monad}) is a lightweight
abstraction that can express the semantics of instrumented lazy evaluation,
suitable for cost analysis.  Based on the ideas of clairvoyant
evaluation~\cite{clairvoyant}, its simplicity is largely due to the absence of
higher-order state commonly associated with laziness.

To model clairvoyant evaluation, we need a monad that can encode all the
following three ingredients: (1)~costs, (2)~nondeterminism, and (3)~failures on
some nondeterministic branches. The clairvoyance monad is the simplest monad
that meets the criterion.

\begin{figure}[t]
\codeplus{example.v}{monad}{language=Coq}
\caption{Core definitions of the clairvoyance monad \cdc{M}.}\label{fig:clairvoyance-monad}
\end{figure}

A computation in the clairvoyance monad, of type \cdc{M a}, nondeterministically
yields a value~\cdc{v : a} after some time \cdc{n}.
A computation is defined as a set of such pairs \cdc{(v, n)}, encoded in Coq as a
predicate \cdc{a -> nat -> Prop}.
The \cdc{ret}urn of the monad yields the given value \cdc{v} with cost 0; it is
a set containing only \cdc{(v, 0)}. The \cdc{bind} of the monad sequences computation by
getting a result \cdc{(x, nx)} from the first operand \cdc{u}, and then feeding the
value \cdc{x} to the continuation \cdc{k}, which then yields another
result \cdc{(y, ny)}. The overall result is that latter value paired with the
total cost \cdc{(y, nx + ny)}.

To track time, \cdc{tick}~\cite{moran1999} is a computation with unit cost. This design follows the same
rationale as \citet{danielsson-08}: explicit ticks make the library lightweight
and flexible to experiment with different cost models.  For a given cost model,
one can ensure that ticks are added consistently by an automatic translation.
We present such a translation in \cref{sec:translation}.

The type of thunks \cdc{T} is structurally an \cdc{option} type. A thunk is
either a known value, under the \cdc{Thunk} constructor, or it is \cdc{Undefined}.
\cdc{Undefined} thunks are placeholders introduced when a computation is
``skipped,'' because its result won't be needed.

For ``laziness,'' we add two operations to create and force thunks.
Intuitively, the \cdc{thunk} function stores a computation of type \cdc{M a}
without evaluating it, and yields a \emph{thunk}: a reference to that stored
computation, of type \cdc{T a}. The \cdc{forcing} function looks up that reference
to evaluate the corresponding computation and passes it to the continuation. This result is
also stored in place of the computation, so that subsequent uses of \cdc{force}
will not recompute the result.

In the clairvoyance model, \cdc{thunk}'s implementation nondeterministically
chooses between (1) running the computation, yielding any one of its results in
a \cdc{Thunk}, and (2) skipping it, yielding an \cdc{Undefined} result at no
cost.  The set of possible outcomes is implemented as a predicate: it accepts
any pair \cdc{(Thunk v, n)} such that \cdc{(v, n)} is accepted by the given
computation \cdc{u} and the pair \cdc{(Undefined, 0)}.

The \cdc{forcing} operation accesses the result stored in a thunk and passes it
to a continuation. If there is indeed a value \cdc{Thunk v}, then \cdc{v} is the
result, and we just pass it to the continuation \cdc{k}. We do not need to add
any costs in this step: we already paid the cost of computing \cdc{v} on the
thunk's creation. If the thunk is \cdc{Undefined}, then the computation fails:
it has no result, as denoted by the empty set. Note that the only way to fail
among the above five combinators is to use \cdc{forcing} and that
\cdc{thunk} is the only way to produce thunks to apply \cdc{forcing} to.
In spite of this underlying potential for failure, computations definable with
these combinators always have at least one successful execution by never
skipping a \cdc{thunk}. In that sense, our combinators adequately model a total
language.

The empty computation \cdc{fun _ _ => False : M a} could also be added to the
core definitions.  In this paper, we will stick to total functional
programming~\cite{turner}.

When programming, we also rely on Coq notations for
a few well-known monadic operations in addition to these core definitions.
The infix notation ``\cdc{>>}'' abbreviates \cdc{bind} with a
constant continuation:
\begin{lstlisting}[language=Coq]
Notation "t >> s" := (bind t (fun _ => s)).
\end{lstlisting}
In the clairvoyance monad, a common idiom is \cdc{tick >> t} to increase the
cost of \cdc{t} by one.

Functions whose arguments are thunks are called \emph{lazy}, in the sense that
their arguments may not always be defined. Otherwise they are \emph{eager}.  Let
us define the following notations, wrapping the monadic \cdc{bind} in more
familiar syntax, akin to do-notation in Haskell and overloaded \cdc{let} in
OCaml.  The infix \cdc{\$!} is named after a standard Haskell operator which
makes a function strict.  For reference, typing rules for these constructs are
given in \cref{fig:notation-typing}.
\begin{lstlisting}[language=Coq]
Notation "let! x  := t in s" := (bind t (fun x => s)).
Notation "let~ xA := t in s" := (bind (thunk t) (fun xA => s)).
Notation "f $! xA" := (forcing xA f).
\end{lstlisting}
We thus view the combinator \cdc{bind} as an ``eager'' \cdc{let!} construct,
where the bound variable \cdc{x} is the result of the computation \cdc{t}.  In
contrast, a composition of \cdc{bind} and \cdc{thunk} provides a ``lazy''
\cdc{let~}, where \cdc{xA} is a thunk for the ``delayed'' computation \cdc{t}.
In this paper, variables of ``lifted'' types \cdc{T a} will be marked with a
suffix \cdc{-A}, to constrast with variables of ``unlifted'' types \cdc{a}.

\begin{figure}
\newcommand\zcdc[1]{\let\par\endgraf\cdc{#1}}
\begin{mathpar}
  \inferrule
    {\Gamma \vdash \hbox{\zcdc{t : M a}} \\
     \Gamma, \hbox{\zcdc{(x : a)}} \vdash \hbox{\zcdc{s : M b}}
    }
    {\Gamma \vdash \hbox{\zcdc{let! x := t in s : M b}}}
    {\hbox{\zcdc{let!}}}

  \inferrule
    {\Gamma \vdash \hbox{\zcdc{t : M a}} \\
     \Gamma, \hbox{\zcdc{(xA : T a)}} \vdash \hbox{\zcdc{s : M b}}
    }
    {\Gamma \vdash \hbox{\zcdc{let~ xA := t in s : M b}}}
    {\hbox{\zcdc{let~}}}

  \inferrule
    {\Gamma \vdash \hbox{\zcdc{f : a -> M b}} \\
     \Gamma \vdash \hbox{\zcdc{xA : T a}}
    }
    {\Gamma \vdash \hbox{\zcdc{f \$! xA : M b}}}
    {\hbox{\zcdc{(\$!)}}}
\end{mathpar}
\caption{Typing rules for \cdc{let!}, \cdc{let~}, and \cdc{\$!}}\label{fig:notation-typing}
\end{figure}

The definition of \cdc{T} has two important features. First, a thunk is merely a
value, not an indirect location in a ``heap'' as it would be in natural
semantics~\cite{launchbury1993}; this is key to the simplicity of our
definitions.
Second, the type constructor \cdc{T} can be used in inductive type declarations
and plays well with the strict positivity condition imposed on
them.~\footnote{\url{https://coq.inria.fr/refman/language/core/inductive.html\#strict-positivity}}
This will be essential in the translation of recursive types
in \cref{sec:translation}.

In the clairvoyance model, it is useful to think of thunks as a way to construct
\emph{approximations}~\cite{scott76}. The type \cdc{T a} ``lifts'' a type
\cdc{a} with an \cdc{Undefined} value which approximates all values of type
\cdc{a}.  Recursive types may contain nested thunks, thus defining rich domains
of approximations. In the monad \cdc{M}, we view a lazy computation as producing
an approximation of some pure, complete result; more precise approximations
are more costly to compute. That structure will be made explicit in
\cref{sec:reasoning}.

\paragraph{Remark}
The monad \cdc{M} coincides with the writer monad
transformer~\citep{monad-transformers} applied to the powerset monad \cdc{_ ->
Prop}. This observation crisply summarizes the orthogonal roles of
nondeterminism and accounting for time in the clairvoyance monad.

\section{Translation}\label{sec:translation}

The clairvoyance monad provides us with an explicit model of laziness.
To reason about the cost of programs in a lazy language---where laziness is
implicit---we make computations explicit by translating them into monadic
programs. We present such a translation in this section. We will then introduce
a framework for reasoning about computations in the clairvoyance monad in the
next section.

Our source language is a total lazy calculus with folds, enabling structural
recursion. In practice, our programs are embedded in Gallina, a pure and total
language which does not prescribe an evaluation strategy---multiple evaluation
strategies are actually available. Hence, Gallina terms have no inherent
notion of cost. Our monadic translation defines a second embedding that
determines a lazy evaluation strategy, allowing us to reason about the cost of
lazy functional programs.

Totality is arguably not a strong limitation for implementing many algorithms:
in the context of complexity analysis, termination is a necessary condition
for defining the cost of an algorithm.

In this section, we formalize the source language and its translation to Gallina.
Our examples of source programs are embedded in Gallina, and we currently
translate them by hand.
This formalization will serve as the basis for a future implementation of a
translation from Haskell.

\subsection{Formal Monadic Translation}\label{sec:semantics}

We define a translation from a simply typed lambda calculus to Gallina using the
monad \cdc{M} and the definitions in \cref{fig:clairvoyance-monad}. The syntax
of the source calculus is summarized in \cref{fig:lambda-syntax}. A primitive
type of lists serves to illustrate how to translate algebraic data types, with
structural recursion modeled by a $\foldsym$ operator.  This calculus is in
A-normal form~\cite{anf} to streamline the monadic translation by confining the
bookkeeping of thunks to $\letsym$ and $\foldsym$.

\begin{figure}[t]
\begin{align*}
\emph{types}\qquad    & \tau  &\ ::=\ & \tau \to \tau \;|\; \listty\;\tau \;|\; \unitty \\
                      & x,y,z & \in\ & \textit{variables} \\
\emph{terms}\qquad    & t,u   &\ ::=\ & x \;|\; \lambda x.\,t \;|\; t\;x \;|\; \lettm{x}{t}{u} \\
                &       &\ \;|\;& \niltm \;|\; \constm\;x\;y \;|\; \foldtm{t}{(\lambda x\, y. t)}{z}
\end{align*}

\begin{mathpar}
  \inferrule
    {\Gamma \vdash x : \listty\;\tau_1 \\
     \Gamma \vdash t_l : \tau_2 \\
     \Gamma,\, y_1 : \tau_1,\, y_2 : \tau_2 \vdash t_n : \tau_2
    }
    {\Gamma \vdash \foldtm{t_l}{(\lambda y_1 y_2. t_n)}{x} : \tau_2}
    {\foldsym}
\end{mathpar}
\caption{Syntax and typing rules for $\foldsym$}
\label{fig:lambda-syntax}
\end{figure}

\newcommand\den[1]{\llbracket#1\rrbracket}

In the translation of types (\cref{fig:type-translation}),
source function types $\tau_1 \to \tau_2$ are translated to
target function types $\hbox{\cdc{T}}\;\den{\tau_1} \to \hbox{\cdc{M}}\;\den{\tau_2}$.
The argument is wrapped in a thunk, so functions may be defined on undefined inputs.
And the result is, of course, a computation.
The type of lists is translated to an inductive type where fields are wrapped in a
thunk \cdc{T}. This type thus represents partially defined lists, which can be seen
as \emph{approximations} of actual lists.

In the translation of terms (\cref{fig:term-translation}),
a well-typed term $t : \tau$ is translated to a Gallina term $\den{t} :
\hbox{\cdc{M}}\;\den{\tau}$.  We pun source variables $x : a$ as target
variables $x : \hbox{\cdc{T}}\;\den{a}$.  A \cdc{tick} is added uniformly in the
interpretation of every non-value construct---this follows \citet{clairvoyant}:
it is assumed that those constructs will be implemented in constant time.  In
our examples, we will simplify ticks further, as discussed in
\cref{subsec:simplify-tick}.

Types guide the design of the term translation.
The source \cdc{Let} corresponds to our lazy \cdc{let~}, creating thunks,
while variable expressions $\den{x} : \hbox{\cdc{M}}\;\den{\tau}$ force the
thunk denoted by the variable $x : \hbox{\cdc{T}}\;\den{\tau}$.

The translation of $\foldsym$ is defined in \cref{fig:foldr-def}.
A \cdc{tick} happens at every recursive call. And recursive calls are
thunked using the \cdc{let~} construct, so that they may remain unevaluated
if the \cdc{c} computation doesn't need them.

Recursion introduces a wrinkle in our translation. Generally, function
arguments $x : a$ are lifted to $x : \hbox{\cdc{T}}\;\den{a}$. However, recursive
definitions in Coq must take an argument whose outer type constructor
is defined using recursion, \cdc{listA a}, unlike the type of thunks \cdc{T (listA a)}.
Thus the translated \cdc{foldrA} is merely a wrapper around the recursive function
\cdc{foldrA'} where most of the work happens.
Moreover, in \cdc{foldrA'}, the subterm \cdc{x2} is forced in
continuation-passing style, using \cdc{forcing} (under the notation \cdc{\$!}),
in order to ensure that the recursive call to \cdc{foldrA'} is syntactically
applied to a subterm of the initial list \cdc{x'}.

\begin{figure}
\begin{minipage}{0.3\textwidth}
\begin{align*}
   \den{\tau_1 \to \tau_2} &= \text{\cdc{T}}\;\den{\tau_1} \to \text{\cdc{M}}\;\den{\tau_2}
\\ \den{\listty\;\tau} &= \text{\cdc{listA}}\;\den{\tau}
\end{align*}
\end{minipage}
\quad
\begin{minipage}{0.5\textwidth}
\codeplus{example.v}{listA}{language=Coq}
\end{minipage}
\caption{Monadic type translation on the left; definition of \cdc{listA} on the right}\label{fig:type-translation}
\end{figure}

\begin{figure}
\begin{align*}
   \den{\lettm{x}{t}{u}} &= \text{\cdc{tick >> let~}\;$x$\;\cdc{:=}\;$\den{t}$\;\cdc{in}\;$\den{u}$}
\\ \den{x} &= \text{\cdc{tick >> force}}\;x
\\ \den{\lambda x.t} &= \text{\cdc{ret (fun}\;$x$\;\cdc{=>}\;$\den{t}$\cdc{)}}
\\ \den{t\;x} &= \text{\cdc{tick >> let! f :=}\;$\den{t}$\;\cdc{in}\;\cdc{f}\;$x$}
\\ \den{\niltm} &= \text{\cdc{ret NilA}}
\\ \den{\constm\;x\;y} &= \text{\cdc{ret (ConsA}\;$x$\;$y$\cdc{)}}
\\ \den{\foldtm{t_l}{(\lambda y_1 y_2. t_n)}{x}} &= \text{\cdc{foldrA}\;$\den{t_l}$\;\cdc{(fun}\;$y_1$\;$y_2$\;\cdc{=>}\;$\den{t_n}$\cdc{)}\;$x$}
\end{align*}
\caption{Monadic term translation}\label{fig:term-translation}
\end{figure}

\begin{figure}
\codeplus{example.v}{foldr}{language=Coq}
\caption{Definition of the \cdc{foldrA} function used in the translation of $\foldsym$}
\label{fig:foldr-def}
\end{figure}

\subsection{Equivalence with Clairvoyant Call-By-Value}\label{subsec:compare-clairvoyant}

Our translation $\den{\cdot}$ is a cost-aware denotational semantics for the simply-typed lambda calculus.
Formally, this semantics $\den{t}$ is parameterized by an
\emph{environment} $\rho$ that maps the free variables of $t$ to semantic
values. The denotation $\den{t}(\rho) : \hbox{\cdc{M}}\;\den{\tau}$ is thus a
set of cost-value pairs $(n, v)$, with $n : \mathbb{N}$ and $v : \den{\tau}$.

We have proved the equivalence between our denotational semantics
and \emph{clairvoyant call-by-value} evaluation, the operational semantics of
laziness introduced by \citeauthor{clairvoyant}, which itself was proved
equivalent to the natural semantics of \citet{launchbury1993}.
Clairvoyant call-by-value evaluation is defined as an inductive relation
$\ccbv{t}{h}{n}{u}{h'}$ expressing that the term $t$ with an
initial \emph{heap} $h$ evaluates, with cost $n$, to a \emph{value term} $u$
and a final heap $h'$.
To state this equivalence, we extend our denotation function $\den{\cdot}$ to these auxiliary syntactic
constructs: a heap $h$, which maps variables to terms, denotes an environment
$\den{h}$, and a (syntactic) value term $u : \tau$, in an environment $\rho$,
denotes a (semantic) value $\den{u}(\rho) : \den{\tau}$.

We can now state an equivalence between our denotational semantics $\den{\cdot}$
and the operational semantics $\Downarrow^{\text{CV}}$:
\begin{theorem}
For any well-typed term $t$ and heap $h$, and for any value-cost pair $(v,n)$,
the following propositions are equivalent.
\begin{enumerate}
  \item $(v,n) \in \den{t}(\den{h})$.
  \item There exists $u$ and $h'$ such that $\ccbv{t}{h}{n}{u}{h'}$ and $v = \den{u}(\den{h'})$.
\end{enumerate}
\end{theorem}
The forward direction, (1) implies (2), states the \emph{adequacy} theorem:
the denotational semantics $\den{\cdot}$ is a subset of the operational
semantics $\Downarrow^{\mathrm{CV}}$. Conversely, the backward direction, (2) implies (1),
states our \emph{soundness} theorem: all evaluations by the operational semantics produce denoted cost-values. Together, these
results prove that our semantics is equivalent to the operational semantics
of \citeauthor{clairvoyant}.

We have formalized both the denotational semantics $\den{\cdot}$ and the
operational semantics $\Downarrow^{\text{CV}}$ and proved the equivalence
theorem in Coq~\citep{forking-paths-artifact}. The proof, which proceeds by
induction on $n$ for adequacy and on derivations of $\Downarrow^{\mathrm{CV}}$
for soundness, is straightforward thanks to the simplicity of the language,
notably excluding general recursion. Most of the work is devoted to relating
mutable heaps $h$ in the operational semantics $\Downarrow^{\mathrm{CV}}$ to
environments $\rho$ in our denotational semantics $\den{\cdot}$.

In addition to an operational semantics $\Downarrow^{\mathrm{CV}}$,
\citet{clairvoyant} also presented a denotational cost semantics, which we can
compare to ours.
First, the cost semantics of \citeauthor{clairvoyant} is defined for an untyped
recursive calculus and our translation is defined for a typed calculus with
folds---guaranteeing termination. However, since the clairvoyance monad is
based on the powerset monad \cdc{_ -> Prop}, we can also define a fixpoint
operator (an example is \cref{fig:fixpoint}):
\[ \hbox{\cdc{Fix : ((a -> M b) -> (a -> M b)) -> (a -> M b)}} \]
Such an operator could be used for the denotational semantics of a general
recursive lazy language, but at the cost of a more complex equivalence theorem.
The issue is that the unfolding lemma \cdc{Fix F <-> F (Fix F)} assumes the
monotonicity of \cdc{F}, adding a significant burden to using that operator.
Without using \cdc{Fix}, we have only modelled a total language, as the source
language we considered above is intended to be a subset of Gallina. Many
algorithms, in the functional programming literature especially, are defined
using various forms of structural recursion, so they can be embedded in our
framework.

\begin{figure}
\codeplus{example.v}{Fix}{language=Coq}
\caption{Possible fixpoint combinator in the monad \cdc{M}}\label{fig:fixpoint}
\end{figure}

Second, the denotations of \citeauthor{clairvoyant} are cost-value pairs that
inhabit a lattice to handle general recursion; they handle
nondeterminism by joining all executions together. However, in the denotation
of \textsc{Let}, the cost of evaluating the binding is discarded if the body of
the \textsc{Let} does not depend strictly on the binding. In comparison, our
semantics models computation using sets of pairs, so the cost of every
nondeterministic path is preserved.

Key to the simplicity of our approach, the core operations of our clairvoyance
monad~(\cref{sec:clairvoyance-monad}) are operations on sets; these operations do not rely
on an abstract lattice structure for values.  In exchange, our semantics is
less well-behaved: \cdc{let~} expressions with unused thunks generate spurious
approximations. We present a dual logic that disregards such uninformative
approximations in \cref{sec:reasoning}.

\subsection{An Example}\label{subsec:translation-example}

We show our translation in action on the example of \cdc{append} and \cdc{take},
illustrating a few pragmatic tweaks to our formalization above.  The source
program with pure functions in \cref{fig:source-example} is translated into the
monadic program in \cref{fig:translated-example}. For simplicity, we retain the
use of fixpoints instead of representing all recursion with \cdc{foldrA}.

These definitions use the \cd{listA} type from \cref{fig:type-translation}.
This type is the corresponding \emph{approximation type} for Gallina's
native \cd{list}, and wraps every field in the thunk type
constructor \cdc{T}.

\begin{figure}[t]
\codeplus{example.v}{trans}{language=Coq}
\caption{The translated code of \cdc{append} and \cdc{take} from the pure version of \cref{fig:source-example}.}\label{fig:translated-example}
\end{figure}

The translation of recursive functions follows a similar structure to the
definition of \cdc{foldrA} in the previous section, since \cdc{append}
and \cdc{take} are in fact specialized list folds (\cdc{foldr}): their
translations are wrappers for the recursive \cdc{append_} and \cdc{take_} where
pattern-matching happens, and the recursive calls are guarded by thunks.

We keep the primitive representation of certain types, such as \cdc{nat} in the
definition of \cdc{take}, instead of using its Peano representation.  The main
reason to do so is that it makes the resulting program simpler by denoting
``primitive'' operations more directly. Although this is generally unsound
for a language with pervasive laziness, this issue could be palliated by using a
strictness analysis to ensure that variables of that type are never instantiated
with $\bot$.
Alternatively, we could consider a source language where both lifted and
unlifted types coexist---Haskell is actually such a language, although unlifted
types are not commonly used because GHC's strictness analysis is often good
enough to enable optimizations.

The \cdc{append} function, of type \cdc{list a -> list a -> list a}, is
translated to its approximate version \cdc{appendA}, of type
\cd{T (listA a) -> T (listA a) -> M (listA b)}.
In other words, the arguments are put under thunks \cdc{T},
and the result is produced by an explicit computation \cdc{M}.
This differs slightly with our formal translation where we simply translated a
term $t : \tau$ to $\den{t} : \hbox{\cdc{M}}\;\den{\tau}$.
We can do away with that outer \cdc{M} because, typically, top-level functions
are values.

Finally, we translate the bodies of the functions. To match the syntax of
\cref{fig:lambda-syntax}, we sequentialize expressions to ANF~\citep{anf} (if they are not already in this form), so that every
computation happens at a \textsc{Let} binding. We then translate the ANF
program to a monadic program, following \cref{fig:term-translation}.

With the translated code, we can now formally analyze its cost. We will show how
we specify its cost and how we reason about that in \cref{sec:reasoning}.

\subsection{Simplifying Ticks}\label{subsec:simplify-tick}
In our examples, we have simplified the translated code further to only
keep a single \cdc{tick} at the head of source function bodies. This incurs a
change to the cost of computations bounded by a multiplicative constant of the
original cost. Considering that those costs are purely abstract quantities
to begin with, this seems an acceptable trade-off to make the translated code
more readable.

This simplification can be broken down in two steps. First, apply the following
rewrite rules to ``float up'' every \cdc{tick} in every subexpression:
\[
  \hbox{\cdc{bind t (fun x => tick >> k x) = (tick >> bind t k)}}
\]
\[
  \hbox{\cdc{thunk (tick >> u) <= (tick >> thunk u)}}
\]
where the inequality \cdc{<=} means in this context that every execution
\cdc{(x, nR)} on the right-hand side corresponds to an execution \cdc{(x, nL)}
on the left-hand side with the same result but a lower or equal cost \cdc{nL <=
nR}.~\footnote{The equality should be interpreted as extensional equality. We
have formally proved these rewrite rules in Coq~\citep{forking-paths-artifact}.}
That rewriting may increase the cost of programs, which is fine since we are
eventually most interested in finding upper bounds on that cost.  Second, once
all ticks are as high in the program as they can be, we replace all consecutive
ticks with a single one. The resulting ``speed-up'' of the computations is
bounded by a constant multiplicative factor equal to the longest chain of ticks
substituted that way.

\section{Formal reasoning}\label{sec:reasoning}

A guiding principle in designing our methodology is to have reasoning rules that
are both \emph{local} and \emph{modular}. By local, we mean that we can reason
about each function independently; and by modular, we mean we can reason about
the whole program by composing the results of reasoning about its parts.

However, in doing so we face a challenge: clairvoyant call-by-value
evaluation is an \emph{over-approximation} of call-by-need evaluation: it
contains nondeterministic branches that would not appear in an actual
call-by-need evaluation.
Therefore, to reason precisely about call-by-need
execution, we need reasoning rules that are \emph{general} enough to contain
many nondeterministic results, but also \emph{selective} enough to prune
nondeterministic branches that contain redundant steps.

We address this challenge with a dual specification
methodology. For \emph{generality}, a \emph{\pessim} specification talks about
the behaviors on all nondeterministic branches.  For \emph{selectiveness},
an \emph{\optim{}} specification describes the behavior of specific
branches.~\footnote{All the theorems discussed in this section have been
formally proved in Coq and they are available in our
artifact~\citep{forking-paths-artifact}.}

\subsection{The \Optim{} and \Pessim{} Specifications}

The definitions of the \pessim{} specification and the \optim{} specification
are shown in \cref{fig:spec-defs}. Both are parameterized by a \emph{specification relation}
\cdc{r : a -> nat -> Prop} which specifies a set of desired values and costs.
A \pessim{} specification states that all nondeterministic branches of the
program \cdc{u} satisfy the relation \cdc{r}.  On the other hand, the
\optim{} specification requires the existence of \emph{at least one}
nondeterministic branch satisfying the relation \cdc{r}.

We use the following notations to denote these two kinds of
specifications:~\footnote{We omit the Coq notation levels in the code.}
\begin{lstlisting}[language=Coq]
Notation " u {{ r }} " := (pessimistic u r).
Notation " u [[ r ]] " := (optimistic  u r).
\end{lstlisting}

\begin{figure}[t]
\codeplus{example.v}{pess/opt}{language=Coq}
\caption{The definitions of the \pessim{} and \optim{} specifications.}\label{fig:spec-defs}
\end{figure}

\subsection{Reasoning Rules}

We can define a set of reasoning rules for the \pessim{} specification and
the \optim{} specification, respectively. For each kind of specification, we
build some reasoning rules for the five basic monadic combinators
(\cdc{ret}, \cdc{bind}, \cdc{tick}, \cdc{thunk}, and \cdc{forcing}) described
in \cref{sec:clairvoyance-monad}. We can then reason about our programs purely
based on these reasoning rules plus a monotonicity rule.

\newcommand\pessimr[2]{#1\ {\colorPessim\{\!\{}\ #2\ {\colorPessim\}\!\}}}
\newcommand\optimr[2]{#1\ {\colorOptim[[}\ #2\ {\colorOptim]]}}

\begin{figure}[t]
\begin{mathpar}
  \inferrule
    {r\ x\ 0}
    {\pessimr{(\texttt{ret}\ x)}{r}}
    {\hbox{\texttt{ret}}}

  \inferrule
    {\pessimr{u}{\lambda x\ n. \pessimr{(k\ x)}{\lambda y\ m. r\ y\ (n + m)}}}
    {\pessimr{(\texttt{bind}\ u\ k)}{r}}
    {\hbox{\texttt{bind}}}

  \inferrule
    {r\ \texttt{tt}\ 1}
    {\pessimr{\texttt{tick}}{r}}
    {\hbox{\texttt{tick}}}

  \inferrule {\pessimr{u}{r} \\ \forall x\ n, r\ x\ n \rightarrow r'\ x\ n}
    {\pessimr{u}{r'}} {\hbox{monotonicity}}

  \inferrule
    {r\ \texttt{Undefined}\ 0 \\
    \pessimr{u}{\lambda x. r\ (\texttt{Thunk}\ x)}}
    {\pessimr{(\texttt{thunk}\ u)}{r}}
    {\hbox{\texttt{thunk}}}

  \inferrule
    {\forall x, t = \texttt{Thunk}\ x \rightarrow \pessimr{(k\ x)}{r}}
    {\pessimr{(k\ \texttt{\$!}\ t)}{r}}
    {\hbox{\texttt{forcing}}}

  \inferrule
    {\pessimr{u}{r} \\ \pessimr{u}{r'}}
    {\pessimr{u}{\lambda x\ n.r\ x\ n\ \wedge\ r'\ x\ n}}
    {\hbox{conjunction}}
\end{mathpar}
\caption{Reasoning rules for \pessim{} specifications.}\label{fig:pessim-reasoning}
\end{figure}

\begin{figure}[t]
\begin{mathpar}
  \inferrule
    {r\ \texttt{Undefined}\ 0 \ \vee\  \optimr{u}{\lambda x. r\ (\texttt{Thunk x})}}
    {\optimr{(\texttt{thunk}\ u)}{r}}
    {\hbox{\texttt{thunk}}}

  \inferrule
    {t = \texttt{Thunk}\ x \\ \optimr{(k\ x)}{r}}
    {\optimr{(k\ \texttt{\$!}\ t)}{r}}
    {\hbox{\texttt{forcing}}}

  \inferrule
    {\pessimr{u}{r} \\ \optimr{u}{r'}}
    {\optimr{u}{\lambda x\ n.r\ x\ n\ \wedge\ r'\ x\ n}}
    {\hbox{conjunction}}
\end{mathpar}
\caption{Reasoning rules for \optim{} specifications. We omit the rules
for \cdc{ret}, \cdc{bind}, and \cdc{tick} as well as the monotonicity rule
because these rules have the same forms as those of the \pessim{}
specifications.}\label{fig:optim-reasoning}
\end{figure}

\Cref{fig:pessim-reasoning} shows the reasoning rules for \pessim{}
specifications. \cdc{ret x} satisfies the \pessim{} specification \cdc{r} if the
result \cdc{x} and the cost 0 are in the set of \cdc{r}. \cdc{bind u k}
satisfies the \pessim{} specification \cdc{r} if all the results of \cdc{u} can
be composed with the continuation \cdc{k} such that all the final results are in
the set of \cdc{r}. A \cdc{tick} satisfies the \pessim{} specification \cdc{r}
if \cdc{tt}~(the only value of the \cdc{unit} data type in Coq) and 1 are in the
set of \cdc{r}. We also need a monotonicity rule to relax the \pessim{}
specification relation and a conjunction rule to combine \pessim{}
specifications.

The term \cdc{thunk u} satisfies the \pessim{} specification \cdc{r} if both
nondeterministic branches forked off from it satisfy the
relation \cdc{r}. The \cdc{forcing} rule requires that its continuation \cdc{k}
satisfies the relation \cdc{r} when applied to the value contained in
a \cdc{Thunk}; in the case that there is no defined value within the thunk (\ie
forcing an \cdc{Undefined}), the \pessim{} specification is vacuously satisfied
because the computation branch fails.

\Cref{fig:optim-reasoning} shows the reasoning rules for \optim{}
specifications. We omit the rules for the \cdc{ret}, \cdc{bind}, and \cdc{tick}
operators and the monotonicity rule here because they have the same form as
those of the \pessim{} specification. There are two ways to give an \optim{}
specification for \cdc{thunk} terms, corresponding to selecting one of the two
different nondeterministic branches that forked off from the \cdc{thunk}. In the
branch where the computation is skipped, we only need to show
that \cdc{Undefined} and 0 are in the relation \cdc{r}. In the branch where the
computation is evaluated, we show that there exists a result in the
computation \cdc{u} such that wrapping it in a \cdc{Thunk} constructor satisfies
the relation \cdc{r}. The \cdc{forcing} rule requires its argument to be a
defined value because forcing an \cdc{Undefined} results in failure. When
reasoning about a program, we need to select the proper \optim{} rule
at \cdc{thunk}s so that forcing an \cdc{Undefined} value never happens.

The conjunction rule for the \optim{} specification is also slightly different
because its premises require both a \pessim{} specification and an \optim{}
specification.

\subsection{Approximations}\label{sec:approx}

Before showing how we use both the \pessim{} and the \optim{} specifications
for reasoning about lazy programs, we need to answer this question: in what
sense does an approximation function implement a pure function?

Recall the approximation types and pure types discussed
in \cref{subsec:translation-example}. We would like to base our specification on
pure types, as this is what we normally write as functional programs. On the
other hand, our implementation in the clairvoyance monad uses approximation
types.

We can connect these approximation and pure types together.  First observe
that we can inject pure types into partial types by thunking each subterm.
We call the result an \emph{exact} approximation because it constructs an
approximation which represents \emph{exactly} the original list.

\begin{lstlisting}[language=Coq]
Definition exact : list a -> listA a.
\end{lstlisting}

We cannot go the opposite way with a function, since approximations generally
contain less information than a full list. Instead, we generalize \cdc{exact}
as a relation \cdc{is_approx xsA xs} between a pure list \cdc{xs} on
the right and any of its approximations on the left. A notation turns
it into an infix operator with syntax inspired by Haskell.

\begin{lstlisting}[language=Coq]
Definition is_approx : listA a -> list a -> Prop.
Infix "`is_approx`" := is_approx.
\end{lstlisting}

Approximations themselves are partially ordered, when the second is at least
as defined as the first. We also use infix notation for this relation.

\begin{lstlisting}[language=Coq]
Definition less_defined : listA a -> listA a -> Prop.
Infix "`less_defined`" := less_defined.
\end{lstlisting}

In our running example using lists, we also simplify things by using the same
type \cdc{a} as the type of elements for pure lists \cdc{list a} and list
approximations \cdc{listA a}.

More generally, we want to overload the function \cdc{exact} and relations \cdc{is_approx} and
\cdc{less_defined}, so that (1) their names can be reused for user-defined data
types, (2) they are automatically lifted through the thunk type constructor \cdc{T}.

Some properties describe and relate these three relations formally. These
relations must be defined and their properties must be proved for every
user-defined approximation type; those propositions and their proofs (which we
omit) follow a common structure, so we conjecture that they can be automated.

The \cdc{less_defined} relation is an order relation.

\label{prop:less-defined-order}
\begin{lstlisting}[language=Coq]
Proposition less_defined_order : Order less_defined.
\end{lstlisting}

The set of approximations for a pure value is downward closed.

\label{prop:approx-down}
\begin{lstlisting}[language=Coq]
Proposition approx_down :
  xsA `less_defined` ysA -> ysA `is_approx` xs -> xsA `is_approx` xs.
\end{lstlisting}

The list \cdc{xsA} is an approximation of \cdc{xs} if and only if \cdc{xsA} is
less defined than the exact approximation of \cdc{xs}.

\label{prop:approx-exact}
\begin{lstlisting}[language=Coq]
Proposition approx_exact : xsA `is_approx` xs <-> xsA `less_defined` (exact xs).
\end{lstlisting}

Exact approximations are maximal elements for the \cdc{less_defined} order.

\label{prop:exact-max}
\begin{lstlisting}[language=Coq]
Proposition exact_max : exact xs `less_defined` xsA -> exact xs = xsA.
\end{lstlisting}

A reference implementation of all the definitions shown in this section as well
as proofs for the above propositions on thunks and lists can be found in our
public artifact~\citep{forking-paths-artifact}.

\subsection{Functional Correctness}\label{sec:fun-correct}

To say that our approximation function implements the pure specification,
we would like two notions of correctness: (1)~a \emph{partial correctness}
notion that requires all the nondeterministic results of the approximation
function to be approximations of the result of the pure function; and (2)~a
\emph{pure correctness} notion that states the existence of a maximal
approximation result that is exactly the pure function's result.

We define the partial correctness of a function using the \pessim{}
specification, and the pure correctness of a function using the \optim{}
specification. For example, the partial and pure specifications
of \cdc{appendA}~(\cref{subsec:translation-example}) are shown
in \cref{fig:fcorrectness}. Given approximations of two input lists \cdc{xs}
and \cdc{ys}, \cdc{appendA} always, \ie pessimistically, yields an approximation
of \cdc{append xs ys}.
On the other hand, \cdc{appendA} optimistically yields exactly the
list \cdc{append xs ys} when applied to the exact approximations of \cdc{xs}
and \cdc{ys}. Both theorems can be proved by an induction over \cdc{xs}.

\begin{figure}[t]
\begin{lstlisting}[language=Coq]
Theorem appendA_correct_partial {a} :
  forall (xs ys : list a) (xsA ysA : T (listA a)),
    xsA `is_approx` xs -> ysA `is_approx` ys ->
    (appendA xsA ysA) {{ fun zsA _ => zsA `is_approx` append xs ys }}.

Theorem appendA_correct_pure {a} :
  forall (xs ys : list a) (xsA ysA : T (listA a)),
    xsA = exact xs -> ysA = exact ys ->
    (appendA xsA ysA) [[ fun zsA _ => zsA = exact (append xs ys) ]].
\end{lstlisting}
\caption{Definitions of \emph{partial} functional correctness and \emph{pure} functional correctness.}\label{fig:fcorrectness}\label{fig:partial-fun-correct}
\end{figure}

\subsection{Cost Specifications}\label{sec:cost-specification}

In this section, we show how we use both the \pessim{} and the \optim{}
specifications for reasoning about computation costs.

Using \cdc{appendA} as our running example, we first start with a \pessim{}
specification.
Since a \pessim{} specification describes all the nondeterministic branches of
a clairvoyant call-by-value evaluation, it might contain spurious branches which
overapproximate call-by-need evaluation too much. Thus, it can only offer a
loose upper bound for the computation cost.
Nevertheless, it is useful in specifying the lower bound,
while we can rely on an \optim{} specification to tighten the bounds.

Taking the \cdc{appendA} function~(\cref{fig:translated-example}) again as our
example, we can give it a \pessim{} specification as follow:
\codeplus{example.v}{appendA_cost_interval}{language=Coq}
The \cdc{xsA} and \cdc{ysA} passed to the \cdc{appendA} function are
approximations of the pure values \cdc{xs} and \cdc{ys}.

The size of approximation lists, defined in \cref{fig:sizex}, is a function
intended purely for reasoning, hence we name it \cdc{sizeX}, with a different
suffix from implementations such as \cdc{appendA}.  It is also parameterized by
the ``size'' of \cdc{NilA}, which is 0 or 1 depending on whether its presence
matters for a given specification.  Here, the extra unit of time accounts for
the final call to \cdc{appendA} which matches on \cdc{NilA}.

\begin{figure}
\begin{lstlisting}[language=Coq]
Fixpoint sizeX {a} (n0 : nat) (xs : T (listA a)) : nat :=
  match xs with
  | Thunk NilA => n0
  | Thunk (ConsA _ xs1) => S (sizeX n0 xs1)
  | Undefined => 0
  end.
\end{lstlisting}
\codeplus{example.v}{isdefined}{language=Coq}

\caption{Definition of \cdc{sizeX}\protect\footnotemark\ and \cdc{is_defined}.}\label{fig:sizex}\label{fig:isdefined}
\end{figure}

\footnotetext{Simplified for clarity. This is actually ill-formed
because the type \cdc{T (listA A)} is not a recursive type
(cf. \cref{sec:semantics}).}

A problem with this specification is that it only gives a range of the
computation costs. During the actual evaluation of the function \cdc{p},
the \cdc{takeA} function would never require more than the first \cdc{n}
elements of \cdc{appendA}'s resulting list, but this specification
of \cdc{appendA} fails to reflect that. We will only be able to compute that the
combined cost has a lower bound of $3$ and an upper bound of $\hbox{\cdc{(sizeX
1 xsA)}} + \hbox{\cdc{n}} + 1$, while in an actual call-by-need evaluation, the
cost would never exceed $2\hbox{\cdc{n}} + 1$.~\footnote{Note that the size
of \cdc{xsA} can be bigger than \cdc{n} if it is required with a higher demand
elsewhere.}

To address this problem, we give an \optim{} specification to \cdc{append}.
A first version states that \cdc{appendA} reaches the lower bound of the
interval in at least one execution.
\codeplus{example.v}{appendA_whnf_cost}{language=Coq}
The execution of \cdc{appendA} which satisfies that specification immediately
discards the computation in the \cdc{let~}, producing only a result in WHNF.

That specification could be strengthened to an equality \cdc{cost = 1}.
However, it is important to remember that results \cdc{(zsA, cost)} of a
clairvoyant computation are formal approximations of the behavior of a lazy
program. \cdc{zsA} is an approximation of the function's result, and \cdc{cost}
is an upper bound on its actual cost. Hence, only upper bounds on \cdc{cost} are
meaningful in \optim{} specifications, while \pessim{} specifications can assert
both lower and upper bounds. For that reason, we leave specifications
of \cdc{cost} as inequalities even though the simple specifications in this
section are technically valid with equalities. Note also that \pessim{} upper
bounds are quite fragile; they can be broken simply by adding spurious thunks in
a program.

Optimistic specifications about a single result such as \cdc{appendA_whnf_cost}
are unhelpful in most proofs, of course. A more expressive way to phrase
\optim{} specifications is to set an arbitrary \emph{demand}, raising the cost
accordingly.

Examining executions of \cdc{appendA} more closely, we can distinguish two
phases, with separate specifications. Before reaching the end of the first list
\cdc{xsA}, \cdc{appendA} computes an approximation of length \cdc{n} in time
\cdc{n}, for any \cdc{n} smaller than the size of \cdc{xsA}.
\codeplus{example.v}{appendA_prefix_cost}{language=Coq}

The natural number \cdc{n} represents an explicit demand on the output of
\cdc{appendA xsA ysA}: we demand an approximation with \cdc{n} constructors
\cdc{ConsA}, costing at most \cdc{n} units of time.

Another specification describes the execution of \cdc{appendA} that reaches
the end of the first list, yielding the most defined result---limited only by
the possible partiality of \cdc{ysA}. As a necessary condition, \cdc{xsA} must
be an exact approximation---modulo the definedness of its elements. Once we
reach the end of the list \cdc{xsA}, the thunk \cdc{ysA} will be forced, so we
require it to be defined, using the \cdc{is_defined} predicate in
\cref{fig:isdefined}. This guarantees that \cdc{zsA} will be defined past the
end of \cdc{xsA}, as specified by assigning a nonzero size to \cdc{NilA} in
applications of \cdc{sizeX}.%
\footnote{The \cdc{(xsA := exact xs)} binding in the following snippet is
desugared into a local definition using \cdc{let}.}
\codeplus{example.v}{appendA_full_cost}{language=Coq}

Natural numbers are not the most precise model of demand on lists: one may also
specify whether and to what extent the elements of the list in the first field
of \cdc{ConsA} constructors should be evaluated. In fact, approximation
types such as \cdc{listA} are the most general way to model demand.
However, when list elements are not explicitly used, a natural number is enough
to formalize the main aspects of complexity analysis for list operations.

\section{Case study: tail recursion}\label{sec:case-study}

We have already demonstrated our methodology on the program described in
\cref{sec:example}. Here, we show how to apply this approach in another context:
reasoning about functions written with tail recursion.~\footnote{All the
theorems discussed in this section have been formally proved in Coq and the
proofs are available in our artifact~\citep{forking-paths-artifact}.} Tail
recursion is a common optimization technique in the context of an eager
semantics. However, it can be a cause of performance degradation if not used
properly under lazy evaluation.

\paragraph{Tail Recursive \cdc{take}}
Consider a tail recursive version of the \cdc{take} function from
\cref{sec:example}, called \cd{take'}.  The key difference is the addition of
an accumulator to its parameters:
\codeplus{example.v}{take'}{language=Coq}
Even though the list must be reversed in the base case, \cdc{take'} is
better in an eager programming language because the
compiler can eliminate stack allocation~\citep{tail-modulo-cons}.

However, the original version is better for lazy evaluation, even if we ignore
the cost of \cdc{rev}. To get an intuition of why, consider the case where we
only demand the WHNF of the resulting list. This variant \cdc{take'} must go
over $n$ elements of the list before it returns the accumulator. In comparison,
the original \cdc{take} can immediately reveal the first element of the
resulting list in any of its pattern matching branches.

\paragraph{A Formal Analysis}
With the help of formal reasoning, we can better understand these functions'
difference from their specifications. In the specifications below, we axiomatize
the cost of \cdc{rev} used in \cdc{take'_} to have a cost of 0~\footnote{The
axiom would not make Coq's logic unsound because we can define such a function
in the clairvoyance monad by not inserting \cdc{tick}s.}. We introduce this
axiom so we can compare only the costs incurred by the recursive calls
on \cdc{take'_} and \cdc{take}. With this set up, let's look at the \pessim{}
specification of \cdc{take'A_}, the version of \cdc{take'_} written in the
clairvoyance monad:~\footnote{For simplicity, we omit the functional correctness
parts of the specifications in this section.}
\codeplus{example.v}{take'A_pessim}{language=Coq}
The \pessim{} specification describes a rather precise cost on all the
nondeterministic branches of \cdc{take'A_}. Furthermore, the cost is purely
decided by the pure values \cdc{n} and \cdc{xs}---the fact that the cost does
not depend on the actual approximations \cdc{zsA} output by \cdc{take'A_} is a
sign that the function may not be making effective use of laziness.

However, to show that \cdc{take} is better than \cdc{take'}, we need to
show that \cdc{take} can cost less than \cdc{take'}. What specification should
we use to show that? One possibility is the \pessim{} specification.
If we take this approach, we can show that the cost of \cdc{takeA}~(\cdc{take}
in the clairvoyance monad) is upper bounded by \cdc{n} and the size of the approximation \cdc{xsA}:
\codeplus{example.v}{takeA_pessim}{language=Coq}
Since approximations are always smaller than their pure values, the cost
shown here is smaller than that of \cdc{take'A_}---\emph{if} such costs indeed
exists. Unfortunately, the \pessim{} specification, which quantifies universally over all
executions, does not guarantee the existence of an execution. In fact, \cdc{take'A}
admits no execution at all if its arguments are not sufficiently defined,
so it satisfies the above specification vacuously.

To show the existence of certain costs, we need an \optim{} specification:
\codeplus{example.v}{takeA_optim}{language=Coq}

The \pessim{} specification of \cdc{take'A_} and the \optim{} specification
of \cdc{takeA} help unveil the key difference between these two: \cdc{take'A_}
does not make effective use of laziness, while the cost of \cdc{takeA} scales
with its demand.

\paragraph{List Reversal}
On the other hand, there are functions like \cdc{rev} that do benefit from
tail recursion. Consider a naive version without tail recursion:
\codeplus{example.v}{rev'}{language=Coq}
And the version with tail recursion:
\begin{lstlisting}[language=Coq]
Fixpoint rev_ {a} (ys : list a) (xs : list a) : list a :=
  match xs with
  | nil => ys
  | x :: xs => rev_ (x :: ys) xs
  end.

Definition rev {a} (xs : list a) : list a := rev_ nil xs.
\end{lstlisting}
One reason that the non-tail-recursive version is worse is that \cdc{append} has
a non-constant cost, which leads \cdc{rev} to have a cost which grows quadratically
in the length of the input list. However, even if we imagine a constant time
version of \cdc{append} (\eg difference lists~\citep{difference-list},
catenable lists~\citep{purely-functional}), this version would not be
better than the tail-recursive one.~\footnote{If we consider the stack usage and
compiler optimizations, the tail-recursive version would be generally more
efficient and does not risk causing stack overflow.} Intuitively, this is
because both versions need to traverse the entire list \cdc{xs} to return the first
element of the resulting list, which is the last element of the input list.

Again we can inspect these two versions of \cdc{rev} formally to better
understand their difference.
Like above, we axiomatize \cdc{append} to have a cost of 0 so that our analysis
only considers the cost incurred by the recursive calls of \cdc{rev_}
and \cdc{rev'}. This simplification makes \cdc{rev'} cost less but will only
strengthen our claim that \cdc{rev'} would not beat \cdc{rev}.

First, we can show a rather specific cost with the \pessim{} specification
of \cdc{revA}:
\codeplus{example.v}{revA_pessim}{language=Coq}
In that specification, the cost is associated with the pure input value,
and is independent of the output value, which suggests that \cdc{revA} does not
make use of laziness. Indeed, this is true: we must iterate over the entire
list \cdc{xs} to get the first element of the resulting list.

We can also prove that \cdc{rev'A} satisfies the following \pessim{}
specification:
\codeplus{example.v}{rev'A_pessim}{language=Coq}
This specification shows that the cost of \cdc{rev'A} also does not depend the
approximations of \cdc{xs}, confirming our intuition that \cdc{rev'A} must also
iterate over the entire list.

\paragraph{Left and Right Folds}
One famous example concerning laziness is the difference between \cdc{foldl}
and \cdc{foldr}. While it seems that the major difference is just their
directions of folding, they actually have rather different costs. For
simplicity, let's only consider these operations on \cdc{list}s. The definitions
of \cdc{foldl} and \cdc{foldr} are shown below:
\codeplus{example.v}{folds}{language=Coq}

A formal analysis of these functions must also consider the cost of the
function \cdc{f} passed into them. For simplicity, let's assume that \cdc{f} has
a cost of only $1$. We can then prove that the translated versions of the above
two functions satisfy the following \pessim{} specification:
\codeplus{example.v}{foldl_pessim}{language=Coq}
\codeplus{example.v}{foldr_pessim}{language=Coq}

The \pessim{} specifications suggest that \cdc{foldrA} makes better use of
laziness because its cost is bounded by the length of approximation \cdc{xsA}.
However, as we have discussed earlier, we need to show that there are indeed
costs lower than the lower bound of \cdc{foldlA} that exists in some
nondeterministic branches of \cdc{foldrA}. For that, we once again need to show
the \optim{} specification of \cdc{foldrA}:
\codeplus{example.v}{foldr_optim1}{language=Coq}
\codeplus{example.v}{foldr_optim2}{language=Coq}
This concludes that, under lazy evaluation, \cdc{foldl} and \cdc{foldr} have the
same worst-case cost, but \cdc{foldr} has a lower cost if the demand is lower.

\section{Discussion}\label{sec:discussion}

\paragraph{Nondeterminism Monads}
The clairvoyance monad is not the only way to model nondeterminism; the list
monad is another possibility which allows some of these ideas to be implemented
without dependent types---necessary for \cdc{exists} and \cdc{eq}. Furthermore,
the list monad makes clairvoyant computations actually executable. We only
require a notion of choice and emptiness for programming; for reasoning, we need
some theory of membership \cdc{member : M a -> a -> Prop}, which defines a monad
morphism into the powerset monad.  This suggests that choosing the powerset
monad as we do here yields the simplest definitions.

\paragraph{Recursive \textsc{Let}}
A recursive \textsc{Let} for data types is yet another challenge.
It allows the construction of certain infinite structures by ``tying the knot'', such
as:
\begin{center}
\cdc{ones = 1 :: ones} \quad\quad and\quad\quad \cdc{fib = 0 :: 1 :: zip_with plus fib (tail fib)}
\end{center}
However, these represent only a small fraction of programs manipulating
infinite structures, and there are many others, which do not rely on
recursive \textsc{Let}, that our reasoning framework can already handle.
For instance, the coinductive versions of \cdc{append} and \cdc{take} are
also actually modeled by our functions translated from the inductive
versions~(\cref{fig:translated-example}), by viewing our inductive \cdc{listA}
also as an approximation type for infinite lists.

In Haskell, \textsc{Let} is always recursive, but it is so rarely useful for
constructing data (as opposed to functions) that the shadowing it introduces is
often considered undesirable, and there is at least one proposal to disable
it.~\footnote{\url{https://github.com/ghc-proposals/ghc-proposals/blob/68164fb2d5a71b62223a8287c0c0390147c0dc2f/proposals/0000-letrec.md}}

\section{Related Work}\label{sec:related-work}

\paragraph{Clairvoyant Evaluation}
Clairvoyant evaluation was first characterized by \citet{clairvoyant}. The main
inspiration for our paper, this work presented an operational semantics for
laziness as an alternative to the natural semantics of \citet{launchbury1993},
as well as a denotational cost semantics, following precursory ideas
by \citet{maraist1995}. We have compared our translation with the semantics
of \citeauthor{clairvoyant} as well as established an equivalence relation
between them in \cref{subsec:compare-clairvoyant}.

\paragraph{Monadic Translation}

\citet{moggi-monad} uses monads to describe computational effects and defines
various translations corresponding to different calling
conventions. \citet{wadler-monad} follows and describes the
translations for the call-by-value and call-by-name semantics, but leaves the
translation for call-by-need as an open problem. \citet{monad-recursive-types}
further adds positive inductive and coinductive types to these
translations.

\citet{malias} proposes a monadic translation that can be used under all
three different calling conventions, generalizing \citet{wadler-monad}, by
defining a function called \cdh{malias} which would be given different meanings
under different
semantics. In particular, \cdc{malias : M a -> M (M a)} is closely related to
our function \cdc{thunk : M a -> M (T a)}: it occupies the
same position in the translation of $\letsym$. The main difference is that,
whereas \cdc{malias} wraps thunks in computations; we expose thunks \cdc{T} as a
separate type.  In fact, we could define \cdc{malias t = bind (thunk t) force}
to hide thunks as well.  However, exposing \cdc{T} is crucial for use
in nested recursive type declarations; using \cdc{M} instead would violate
Coq's strict positivity condition. Furthermore, in \citet{malias}, \cdc{M} is
defined using Haskell's \cdc{ST} monad, relying on a polymorphic interface of
mutable references to represent the heap. We believe that it would be challenging to
emulate such an interface in Coq.

\paragraph{Computation Cost and Laziness}

There is much work on reasoning formally about computation costs.
For example, \citet{danielsson-08, lambda-amor, raml, implicit-complexity,
resource-bound-certification, timl} study intrinsic approaches to formal cost
analysis. Our work is in the extrinsic context of an interactive proof assistant.

On the extrinsic side, \citet{amortized-union-find, a-fistful-of-dollars} use
separation logic for reasoning about computation costs under call-by-value
evaluation using amortized analysis. Compared to these works, our goal of
reasoning about lazy pure functional programs does not require separation
logic.
Cost specifications could be made more modular by hiding
implementation-specific constant factors and formulating costs in asymptotic
terms. Works on formalizing asymptotic complexity
include \citet{asymptotic-complexity, asymptotic-proof-assistant,cutler20}.

\Citet{danielsson-08, liquidate} reason about lazy functional programs in a
monadic syntax annotated with ticks. An issue in both works is that they require
an explicit notion of laziness to model sharing: for example, in practice, a
list that is evaluated once will not be evaluated again under lazy
evaluation. To avoid a ``double counting'' of the cost in a thunk, a \cdc{pay :
nat -> M a -> M (M a)} combinator with an explicit representation of cost must
be annotated in the code~\footnote{The \cdc{pay} combinator is also an annotated
version of \cdc{malias : M a -> M (M a)}~\cite{malias}.}. This prevents both
works to be fully extrinsic in reasoning about laziness. With the clairvoyance
monad, thunks are either paid for or discarded immediately, so it is impossible
to count the cost of a thunk twice. This enables us to translate pure lazy
functions mechanically to monadic programs, and our proofs are completely
extrinsic.

On the automated reasoning side, \citet{contract-based-lazy} verify a purely
functional subset of Scala by translating higher-order functions to first-order
programs via defunctionalization. They also model memoization by encoding the cache
as an expression that changes during the execution of the program.

For testing lazy functions in Haskell, Sloth is a tool that automatically
generates test cases to check if a function is ``unnecessarily
strict''~\citep{sloth}.  This tool relies on a ``less-strict'' ordering of
functions. One function is less-strict than another when, given the same input,
its result is less defined~\citep{less-strict}.

\Citet{foner2018} develop a library that generates random demands on the output
of a function and instruments inputs to record induced demand. Demands take the
form of approximations whose structure is also derived from pure data types.

\paragraph{Haskell}
Although we only discuss Coq here, Haskell is also a potential target of our approach.

The \hstocoq tool automatically translates Haskell programs to
Gallina---Coq's specification language---using shallow
embeddings~\citep{hs-to-coq}. It has been used for verifying a significant
portion of Haskell's \cdc{containers} library~\citep{containers} and one part of
GHC~\citep{verify-ghc}. However, \hstocoq's pure translation cannot be
used for cost analysis so existing work using this tool has been restricted to
functional correctness.

\Citet{haskell-monadic-translation} and \citet{one-monad} respectively translate
Haskell to monadic embeddings in Agda and Coq, based on the call-by-name
translation by \citet{moggi-monad}. This is enough to model Haskell's
partiality, but not its lazy cost semantics.

\LiquidHaskell augments Haskell with refinement
types~\citep{liquidhaskell-dissertation} to enable formal verification, and it
has been applied to cost analysis~\citep{liquidate}. The major difference is
that our work does not require an explicit notion of laziness, as discussed
earlier in this section. Furthermore, \citet{liquidate} verify Haskell programs
written explicitly in the tick monad; to analyze arbitrary Haskell programs,
some monadic translation is necessary.

\paragraph{Nondeterminism and Dual Logics}

Our optimistic and pessimistic specifications are examples of predicate
transformer semantics. They date back to \citet{dijkstra1975}, forming the
basis of much work on the verification of effectful programs in type
theory~\citep{htt,hoare-st,pt-semantics,dijkstra-monad}.
Our predicate transformer semantics are two conventional effect
observations~\cite{maillard2019} from the clairvoyance monad---a variant of the
powerset monad---to the specification monads respectively for angelic and
demonic nondeterminism.

The duality between \pessim{} and \optim{} specifications is also the duality of Hoare
logic~\citep{hoare-logic} and reverse Hoare
logic~\cite{rev-hoare,incorrectness}.  Those logics use sets of states to
approximate program behavior.  In Hoare logic, the postcondition
over-approximates the set of reachable states; in reverse Hoare logic, the
postcondition under-approximates the set of reachable states.
Here, we show that abstractions for angelic and demonic nondeterminism
give rise, rather simply, to logics of over- and under-approximations
of time consumption.
The notion of approximation underlying our logics is formally defined as
follows: a set of cost-value pairs $A$ underapproximates a set of pairs $B$ if,
for every $(v,c) \in A$, there exists $(w,d) \in B$ which ``costs less and is
more defined'', \ie such that $d \le c$ and $v \le w$.
Thus, sets of states are ordered by inclusion in Hoare logic, whereas sets of
cost-value pairs follow a more elaborate order structure in our dual logic,
based on the view that those pairs themselves are approximations of the actual
behavior of lazy programs.

\section{Conclusion and future work}\label{sec:conclusion}

In this paper, we present a novel and simple framework for formally reasoning
about costs of lazy functional programs. The framework is based on a new model
of lazy evaluation: clairvoyant call-by-value~\citep{clairvoyant}, which makes
use of nondeterminism to avoid modeling mutable higher-order state in classic
models of laziness~\citep{launchbury1993}.

Our framework includes a simple clairvoyance monad and a translation from a
typed calculus to programs in this monad that captures the semantics of
clairvoyant call-by-value. Compared with the denotational semantics
of \citeauthor{clairvoyant}, our translation deals with typed programs, does not
rely on domain theory, and accounts for the cost of every nondeterministic
execution.
We also develop dual logics \emph{over-} and \emph{under-approximations} similar to
those of \citet{hoare-logic,rev-hoare,incorrectness} that enable local and
modular formal reasoning of computation costs. We show the effectiveness of our
approach via several small case studies.

In future work, we would like to apply this methodology to existing programs
written in Haskell. In particular, we would like to augment tools
like \hstocoq~\citep{hs-to-coq} so that they can automatically translate Haskell
programs to the clairvoyance monad and explore techniques in Coq that can be
used to automate reasoning in those logics.

\begin{acks}

We thank Koen Claessen, who recommended to one author the work
of \citet{clairvoyant} during ICFP'20, and Deepak Garg, who recommended to
another author the work of \citet{danielsson-08} during POPL'21. We thank the
anonymous reviewers of ICFP'21, whose feedback helped improve the paper. We also
thank our shepherd Leonidas Lampropoulos.

This material is based upon work supported by
the \grantsponsor{GS100000001}{National Science
Foundation}{http://dx.doi.org/10.13039/100000001} under Grant
No.~\grantnum{GS100000001}{1521539}, and Grant
No.~\grantnum{GS100000001}{2006535}.  Any opinions, findings, and conclusions or
recommendations expressed in this material are those of the authors and do not
necessarily reflect the views of the National Science Foundation.
\end{acks}

\bibliography{ref}

\end{document}